\documentclass[showpacs,nofootinbib,preprintnumbers,amsmath,amssymb,preprint,aps]{revtex4}
\usepackage[pdftex]{graphicx}
\usepackage{amsfonts}
\usepackage{bm}
\usepackage{amsmath}
\usepackage{tensor}
\usepackage{amssymb}
\usepackage{mathtools}
\usepackage{color}
\usepackage[caption=false]{subfig}
\usepackage[all]{xy}

\newtheorem{theorem}{Theorem}[section]

\begin{document}

\title{Singularity theorems in Schwarzschild spacetimes}

\author{Servando V. Serdio$^1$, Hernando  Quevedo$^{1,2,3}$}
\email{svserdio@correo.nucleares.unam.mx, quevedo@nucleares.unam.mx}
\affiliation{
$^1$Instituto de Ciencias Nucleares, Universidad Nacional Aut\'onoma de M\'exico,
 AP 70543, M\'exico, DF 04510, Mexico\\
$^2$Dipartimento di Fisica and ICRANet, Universit\`a di Roma ``La Sapienza",  I-00185 Roma, Italy\\
$^3$Institute of Experimental and Theoretical Physics, Al-Farabi Kazakh National University, Almaty, Kazakhstan}

\date{\today}

\begin{abstract}

We present a review of the two prominent singularity theorems due to Penrose and Hawking, as well as their physical interpretation. Their usage is discussed in detail for the Schwarzschild spacetime with positive and negative mass. First, we present a detailed mathematical proof to formally guarantee the existence of a singularity of geodesic incompleteness for the case of positive mass. Second, we discuss the applicability of the mathematical tools used by the theorems in the negative mass case. The physical implications of the validity or inconsistency of the hypotheses of such theorems on the latter case, are also exhibited. As far as this analysis is concerned, some clues are produced regarding future research that could result in general properties for naked singularities.


\end{abstract}


\maketitle


\section{Introduction}
\label{sec:int}

The theory of singularities in general relativity represents an enlightening theoretical development that, despite being around for about 50 years by now, keeps captivating the attention of beginners and specialists. 
Nobody (even without any training in physics) should doubt its profound significance and implications to our understanding of the Universe. 
The discovery of singularities, as an inevitable (and generic) consequence of one of the most successful physical theories in history, rocked our minds to their core. The experimental astronomical observations to date, together with a vastly tested and many times corroborated theory like general relativity, suggest that: (a) a long (but finite) time ago, the stuff that constitute everything around us might have suddenly started to exists under a set of rules (e.g. the theory itself) that seem to be oblivious of that initial event, and (b) a similar (but time reversed) phenomenon seems to take place when enough energy-matter is brought together inside a bounded spatial region. Such a content of matter (or at least part of it) would collapse on itself until it reaches such an extreme state that its existence comes to an end (i.e. it ceases to exist within the framework of the theory, and cannot be described any longer by means of it)

Whether or not one is willing to accept a sudden ``creation'' or an inevitable final fate as the ultimate consequences of the singularity theorems, they surely state without a doubt a set of circumstances under which one cannot rely anymore on the theory to make predictions. As if general relativity where not astounding enough, certain cases in which its validity cannot be assumed any longer are plainly exposed by the theory itself via the theorems in question.

With such profound implications, and as it should be the case, a huge variety of results have been motivated by the singularity theorems after the first one of their kind (due to Penrose) was published in 1965 \cite{Penrose65}. An extensive and illustrative summary of the antecedents, concepts and consequences of that particular theorem can be found in \cite{SenovillGarfinkl}. Specifically speaking, the novelty of such theorem (and the other classical one that succeeded it, due to Hawking and Penrose \cite{HawkPenr70}) resides in its use of an original concept called a {\it closed trapped surface} (that captures the idea of an inevitable and instantaneous total confinement scenario for an enclosed energy-matter distribution) which, together with other geometric properties of a given spacetime (like an ``energy'' condition and some sort of global causality), tends to favour the formation of {\it geodesic incompleteness} within such spacetime (being this the very same concept representing the aforementioned notion of sudden creation/disappearance of particles, also presented for the first time by Penrose).

It is not an overstatement to say that all the research that have been motivated by the first theorems is vast and can easily lead to bewilderment. Despite the existence of works like \cite{SenovillGarfinkl, Senovill1, Hawking} that can be used as a first guide to the nowadays world of singularities, as soon as one steps outside of them one immediately has to face a menagerie of results that (in one way or another) were inspired by the three notions/concepts previously mentioned. In order to briefly give some particular examples related to the main conclusions derived from the following sections (that in no way constitute an exhaustive list, but that are intended to broaden in here the outlook of their corresponding subjects), three categories (most likely intertwined) can be made.   

First, the concept of a trapped surface has been a very prolific one as it can be generalized in several ways thanks to its possible characterization via the mean curvature vector field \cite{SenovillGarfinkl}. By virtue of this vector field, in particular, closed trapped submanifolds of arbitrary co-dimension can be defined in order to generalize the Hawking-Penrose theorem \cite{GallowaySenovill}. In a similar manner, the Penrose theorem can be generalized to include surfaces to some extent ``less trapped'' (like marginally outer trapped surfaces, for example),
and to approach the possible formation of singularities from an initial data set (i.e. the so called initial data singularity theorems) \cite{EichmairGallowayPollack}. Additionally, and with respect to more recent incursions of the trapped surface concept into the realm of black hole physics, the notion of the smallest trapped region in a spacetime can be used to define the boundary of black holes that do not necessarily belong to stationary spacetimes \cite{BengtssonSenovill}. By virtue of the present work, the notion of a trapped surface could even be used to hint the existence of (curvature) singularities, despite those submanifolds not representing total confinement.

Second, the earliest attempts to address the natural question of whether or not additional information about a singularity of geodesic incompleteness could be extracted from the theory (apart from their mere existence), were promptly made after the first singularity theorems were proven. From theorems stating conditions under which a spacetime could be extended to include a singularity (in a smoothly enough manner) \cite{Clarke1, ClarkeSchmidt, Clarke1a, Clarkebook}, to theorems and results categorizing singularities by the way in which certain scalar magnitudes diverge when traversing the curves leading to them \cite{Clarke3, Clarke4, Thorpe1, KannarRacz}, the issue has shown to be a complicated one to resolve (most likely because, in a strict sense, singularities are not part of the spacetime ``possessing'' them and cannot be analyzed like any other geometric object). As far as the present work is concerned, some restrictions can be imposed on the rate at which the stress tidal tensor (with respect to a parallelly propagated orthonormal/pseudo-orthonormal basis) grows towards a singularity of geodesic incompleteness \cite{KannarRacz}. As exposed in the following analysis of the Schwarzschild spacetime, not every component of such tensor is restrained to diverge with the rate in question towards singularities in null incomplete geodesics. As it turns out, the use of such decomposition for the tidal stress tensor might even provide information regarding the repulsive or attractive nature of a singularity. This feature could be used to shed some light into the soundness of new ideas intended to precisely define repulsive gravitational effects \cite{GutierrzQueved}.

Third, the profound conceptual (and philosophical) repercussions of the theorems have always been largely based on the seeming physical feasibility of their hypotheses. Nevertheless, with only local experimental evidence at our disposal, it is only natural to explore the possibility of weakening or even changing/improving some of those assumptions. Excellent expositions of the ideas behind some efforts on the matter, and the results obtained concerning the so called rigidity theorems, can be found in \cite{Beemetal, CostaeSilvaFlores, GallowayVega014}. Although important generalizations of the classical theorems have been accomplished in relation to, for example, the weakening of requirements like smoothness or global causality \cite{GrafGrantKuzinger, Maedaishibashi}, it also has being possible to relax the boundary/initial condition of the theorems (appearing in the form of trapped submanifolds). As mentioned earlier, the use of the mean curvature vector field allows for ``lesser trapped'' scenarios to be considered, and singularity theorems have been shown to hold for spacetimes possessing marginally and marginally outer trapped surfaces \cite{CostaeSilva012}. One of the main results of the following exposition consists in showing an example of another way in which a trapped surface (that is not closed nor it represents inevitable confinement) could be used to hint the presence of singularities.

Finally, it goes without saying that the Schwarzschild spacetime has always played an invaluable role in the development of the general theory of relativity and its experimental corroboration. Since every introductory relativity book makes use of such solution to derive classical tests of the theory, and to plainly show some illustrative effects, its significance is beyond dispute. Regarding the impact of studies exploiting spherical symmetry to the theory of singular spacetimes, it is worth to recall that from the very first theoretical indication (to more recent ones) of the (possible) production of singularities by gravitational collapse, an exterior vacuum Schwarzschild spacetime has always been matched to another collapsing interior solution \cite{OppnheimrSnydr, Joshibook}. Additionally, such level of symmetry has also been used, among many other things, to understand the global properties of closed trapped surfaces \cite{WaldIyer91}, to attempt to define the boundary of non-stationary black holes \cite{BengtssonSenovill}, and to rule out low classes of extensions with which one could otherwise attach a singularity to its spacetime \cite{Sbierski018}. Within this context, the present article exploits once again the analytical manageability of the Schwarzschild solution to exhibit specific geometrical properties that could hopefully render new topics for singularity research.

The contents of the next exposition have the following structure. In section \ref{sec:roST}, the theorems of Penrose and Hawking-Penrose are presented, along with physically explanatory interpretations for each one of their hypotheses. Section \ref{sec:SchSPpm} contains a precise definition of the innermost Schwarzschild vacuum spacetime with positive mass (i.e. the black hole region of the full Schwarzschild solution, viewed as a manifold on its own), followed by a detailed proof (split into several subsections) of the fulfilment by such spacetime of Penrose's theorem hypotheses. In section \ref{sec:SchSPnm}, the main differences between the cases of positive and negative mass are pointed out in order to precisely define the corresponding spacetime for the latter. In the form of several subsections, the fulfilment or infringement by this second spacetime of the hypotheses of both theorems (associated with compact gravitational sources), are analyzed. Special attention is given to the physical causes and/or implications of the obtained results. Lastly, some concluding remarks are given in section \ref{sec:conclus}. Natural units with $G=c=1$ are used throughout this paper.


\section{Review of two singularity theorems}
\label{sec:roST}

The two singularity theorems used in the following  sections are due to Penrose and Hawking \cite{HawkElli} and nowadays can be regarded as classical. Despite them being well known and understood by now, it seems hard to find in the literature a concise and intuitive way of understanding the physical meaning of their hBecause of this, and taking into account the fact that they can be found in a variety of references, here they will be simply stated right away, followed only by their physical interpretation in the context of a particular spacetime structure. For the sake of completeness,   it should be reminded here that the concept of a {\it spacetime} simply refers to the collection of all possible positions and times (i.e. {\it events}) in a space, together with the capacity to measure temporal and spacial distances. Moreover, it should also be kept in mind that in these theorems a spacetime always refers to a 4-dimensional differential manifold without a boundary that is paracompact, connected, Hausdorff, and which is equipped with a (sufficiently well behaved) Lorentzian metric.

\begin{theorem}[Penrose (1965)]
\label{penrose65}
Let $(M,g_{\alpha\beta})$ be a spacetime for which the following conditions are met: \textup{($1$)} It is connected. \textup{($2$)} It is globally hyperbolic with a non-compact Cauchy surface. \textup{($3$)} At each one of its points the inequality $R_{\alpha\beta}k^{\alpha}k^{\beta} \geq 0$ is satisfied by every null vector $k^{\mu}$. \textup{($4$)} It contains a closed trapped surface. Then $(M,g_{\alpha\beta})$ contains at least one null geodesic that is incomplete
\end{theorem}

{\bf Interpretation.} Given a spacetime such that: ($1$) There do not exist isolated regions within it. ($2$) By knowing the initial conditions in a spacelike unbounded hypersurface, it is possible (in principle) to determine the complete physical state at every other event. ($3$) The gravitational interaction felt by light waves due to an energy-matter distribution is always attractive. ($4$) There exists a spacelike 2-surface, that closes on itself, for which the two families of light rays departing orthogonally from it have instantaneously decreasing area wavefronts \cite{MarsSenovilla}. Then there exists at least one light ray trajectory that cannot be extended, and which ends up after a finite extent.

\begin{theorem}[Hawking-Penrose (1970)]
\label{hawkpenr70}
Let $(M,g_{\alpha\beta})$ be a spacetime for which the following conditions are met: \textup{($1$)} For every causal vector $v^{\mu}$, the inequality $R_{\alpha\beta} v^{\alpha} v^{\beta} \geq 0$ holds. \textup{(2)} There do not exist closed timelike curves within $M$. \textup{(3)} Every causal and inextensible geodesic possesses a point at which its tangent vector $k^{\mu}$ satisfies the \textup{(}so called\textup{)} generic condition $k_{[ \alpha}R_{\beta ] \gamma\delta [\mu}k_{\nu ]}k^{\gamma}k^{\delta} \neq 0$. \textup{(4)} There exists within $M$ at least one of the  following subsets: \textup{(a)} A compact and non-chronological set without an edge. \textup{(b)} A closed trapped surface. \textup{(c)} A \textup{(}trapped\textup{)} point $p$ such that the expansion of the family of null geodesics emanating from it into the future \textup{(}or past\textup{)} always becomes negative along each one of these geodesics. Then $(M,g_{\alpha\beta})$ contains at least one causal geodesic that is incomplete.
\end{theorem}

{\bf Interpretation.} Given a spacetime such that: ($1$) The gravitational interaction due to an energy-matter distribution is always attractive for every collection of massive particles and light waves. ($2$) No massive particle can have any influence on its own past. ($3$) Every freely falling massive object experiments, at some moment in its history, tidal forces within it. Also, along each light ray trajectory there exists a point such that, on accelerating a massive object to near-light velocities in its direction, the tidal forces across its points become a greater issue to overcome than its energy increase \cite{Beemetal}. ($4$) There exists at least one of the following sets: (a) A subset with no endpoints, despite being bounded, whose events cannot influence their own past. (b) A closed trapped surface. (c) A point for which, along every direction, the light waves emanating from it into the future always suffer a contraction at some moment. Then there exists at least one freely falling massive particle trajectory, or a light ray trajectory, that cannot be extended, and which ends up after a finite extent.

\bigbreak
For more details on the physical interpretation of these singularity theorems, see \cite{HawkElli, Waldbook, Krielebook}.


\section{Schwarzschild spacetime with positive mass}
\label{sec:SchSPpm}

It is a common practice in the literature to work out a formal proof of the previous theorems and then simply state the validity of their hypothesis for a given spacetime. In the case of the Schwarzschild solution, the Penrose diagram for its maximal (Kruskal-Szekeres) extension can be used to visualize the applicability of theorem \ref{penrose65} to it. Nevertheless, it does not seem to exist so far a complete and detailed treatment addressing this fact. Even if its confirmation is considered as straightforward and trivial as things can get, carrying its details out definitely sheds some light into the key features that must be taken into account when dealing with the theorems. It also opens up a window into the physical properties that can be learned about a spacetime, via the mathematical tools used in the theorems. Because of all of these reasons, special attention is given next to the definition of the spacetime itself.


\subsection{The Schwarzschild spacetime}
\label{subsec:SchwSpTi}

The spacetime to be considered $(M_{Sch},g_{\alpha\beta})$ consists of only the inner region $(r<2M)$ of the (so called) exterior Schwarzschild solution. Specifically speaking, the manifold is given by the set of points $p \in M_{Sch} \equiv \mathcal{O}$, for which a single chart $\{\left( \mathcal{O}, \Psi\right)\}$ is defined by means of the map
	\begin{equation*}
		 p \xmapsto[\Psi]{\phantom{........}} \left( t(p), x(p), y(p), z(p) \right) \in \mathbb{R}  \times \mathcal{B}_{o}(2M),
	\end{equation*}
where $M>0$ and $\mathcal{B}_{o}(2M) \equiv \left\lbrace (x,y,z)\in \mathbb{R}^3 \;|\; 0 < x^2+y^2+z^2 < 4M^2 \right\rbrace$. The metric $g_{\alpha\beta}$ is introduced in the form (with $x_{(1)}\equiv x$, $x_{(2)}\equiv y$ and $x_{(3)}\equiv z$)
	\begin{equation}
	\label{Schwlinelemcarts}
		\begin{split}		
			ds^{2}=-\left( 1 - \frac{2M}{\sqrt{x_{(i)}x_{(j)}\delta^{ij}}} \right)dt^2 &+ dx_{(i)}dx_{(j)}\delta^{ij} \\
			 &+ \frac{2M}{\sqrt{x_{(i)}x_{(j)}\delta^{ij}}-2M}\left( \delta^{km}\delta^{ln}x_{(k)}x_{(l)}dx_{(m)}dx_{(n)} \right).
		\end{split}
	\end{equation}
	
	\begin{figure}
		\includegraphics[scale=0.35]{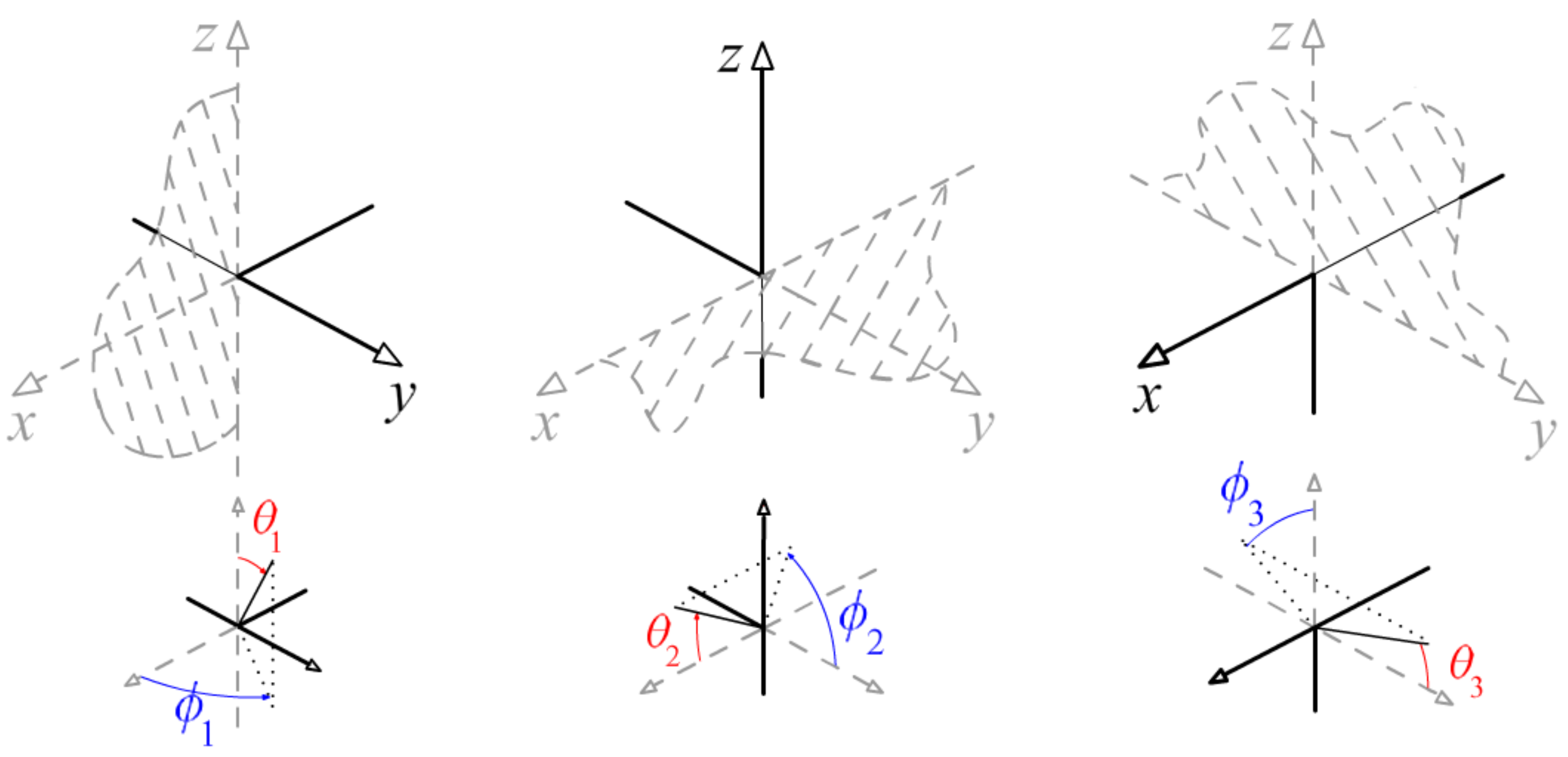}
		\caption{\label{graph.2ndatlas} Three (well defined) different spherical open charts are required in order to completely cover $\mathbb{R}^{3} \setminus \{0\}$.} 
	\end{figure}	

In order to relate this expression to the usual one in spherical coordinates (but in an inconsistency free manner), three additional charts $\{\left( \mathcal{O}_{i}, \Psi_{i}\right)\}^{3}_{i=1}$ must be taken into consideration. The most common transformation between (flat) spherical and cartesian coordinates are used to define these charts. The only difference between each chart being the fact that the vertical direction is taken to coincide with a different cartesian coordinate axis fixed in the background (Fig. \ref{graph.2ndatlas}). For example, the second one of these charts corresponds to the case $y = r \sin\vartheta_2 \cos\varphi_2$, $z = r \sin\vartheta_2 \sin\varphi_2$ and $x = r \cos\vartheta_2$, with $(\vartheta_2,\varphi_2, r) \in (0,\pi) \times (0,2\pi) \times (0,2M)$. It is then a matter of simple algebra to reduce the previous line element to the form
	\begin{equation}
	\label{Schlinelemnt}
		ds^{2}= \frac{2M-r}{r}dt^{2} + r^{2}\left( d\vartheta^2_i + \sin^2\vartheta_i d\varphi^2_i \right) - \frac{r}{2M-r}dr^{2}.
	\end{equation}

Despite the fourth coordinate being timelike in this spacetime, it will remain being designated by the symbol $r$ (i.e. the conventional spherical radial character) because of the way it relates to the cartesian-like coordinates used in the first place. For this same reason, it will also be referred to as the {\it radial} coordinate of the atlas in question. Furthermore, its global nature (i.e. the fact that it takes a certain value for a point $p \in M_{Sch}$, regardless of the neighbourhood  $\mathcal{O}_{i}$ containing it) can be exploited to establish a temporal orientation for the whole $(M_{Sch},g_{\alpha\beta})$. More specifically, every tangent vector $X^\mu$ whose components $(X^{\mu}) = (X^t, X^{\vartheta_i}, X^{\varphi_i}, X^r)$ satisfy the condition $X^r<0$, will be said to be {\it future directed}.

As can be anticipated from the way a spacetime is handled by the singularity theorems of section \ref{sec:roST}, the last construction comprises all of the basic structure required to corroborate whether the hypotheses of the theorems are satisfied or not by $(M_{Sch},g_{\alpha\beta})$. The content of the following subsections include some formal arguments regarding the applicability of theorem \ref{penrose65} to this spacetime.


\subsection{Connectedness}
\label{subsec:cnnctdnss}

Every $\mathcal{O}_i$ is clearly connected since its image, under the homeomorphism $\Psi_i$, is the set $\mathbb{R} \times (0,\pi) \times (0,2\pi) \times (0,2M)$. The fact that no two of the regions $\mathcal{O}_i$ are disjoint, guarantees the connectedness of the whole $M_{Sch}= \cup^{3}_{i=1}\mathcal{O}_{i}$ \cite{Engelkingbook}.


\subsection{Closeness and non-compactness of a hypersurface $\boldsymbol{S_{r_0}}$}
\label{subsec:clsnssnncmpctnss}

By means of the following subsets (with $0 < r_0 < 2M$)
	\begin{equation}
	\label{Cauchsurfc}
		\begin{gathered}
			S_{r_{0}} \equiv \{ p\in M_{Sch} | r(p) = r_0  \}, \quad S_{r<r_{0}} \equiv \{ p\in M_{Sch} | 0 < r(p) < r_0  \} \\
			 \quad \text{and} \qquad S_{r>r_{0}} \equiv \{ p\in M_{Sch} | r_0 < r(p) < 2M  \},
		\end{gathered}
	\end{equation}
a disjoint partition of the manifold $M_{Sch}$ is made. Since $\Psi_{i}(S_{r_0} \cap \mathcal{O}_i)$ is closed in $\mathbb{R} \times (0,\pi) \times (0,2\pi) \times (0,2M)$, it also happens that $S_{r_0} \cap \mathcal{O}_i$ is closed in $\mathcal{O}_i$. If $\overline{A}$ represents the closure of $A \subset M_{Sch}$, it is clear that $S_{r_{0}}\cap\mathcal{O}_i=\overline{ S_{r_{0}}\cap\mathcal{O}_i } \cap \mathcal{O}_i$ and $\overline{S_{r_{0}}}\cap\mathcal{O}_i=\overline{\overline{S_{r_{0}}}\cap\mathcal{O}_i} \cap\mathcal{O}_i$. Because of each $\mathcal{O}_i$ being an open subset of $M_{Sch}$, it is also the case that $\overline{\overline{S_{r_{0}}}\cap\mathcal{O}_i} = \overline{ S_{r_{0}} \cap\mathcal{O}_i}$ \cite{Engelkingbook}. From all the previous statements it is straightforward to obtain $\overline{S_{r_0}} = S_{r_0}$.

On the other hand, $S_{r_0}$ is clearly a Hausdorff topological space (with respect to its induced topology) that can be covered by the collection of open sets $\mathcal{U}_n \subset M_{Sch}$ ($n \in \mathbb{N}$), defined by $\Psi( \mathcal{U}_n \cap \mathcal{O}) \equiv (-n,n) \times [\mathcal{B}_{o} (r_0 + \varepsilon) \setminus \mathcal{B}_{o}(r_0 - \varepsilon)]$ (with $0 < \varepsilon < r_0$). Since no finite subcover for $S_{r_0}$ can be extracted from $\{ \mathcal{U}_n \}_{n \in \mathbb{N}}$, it follows that this hypersurface is not compact.

\bigbreak
It is intuitively clear from the (by now standard) $r-t$ diagram of the radial null geodesics, that $S_{r_0}$ is a good candidate for a Cauchy surface. The purpose of the following four subsections is to formally corroborate that this is indeed the case.


\subsection{Achronality of $\boldsymbol{S_{r_0}}$}
\label{subsec:acronlSr0}

Given a smooth timelike curve $\gamma(\tau)$, parameterized by its arc lenght $\tau$, its tangent vector components $(\dot{t}, \dot{\vartheta}_i, \dot{\varphi}_i, \dot{r})$ (with $\dot{a} \equiv da/d\tau$) satisfy the relation 
	\begin{equation*}
		-1 = \frac{2M-r}{r}\dot{t}^2 + r^{2} \left( \dot{\vartheta}^2_i + sen^2\vartheta_i \dot{\varphi}^2_i \right) - \frac{r}{2M-r}\dot{r}^2.
	\end{equation*}
	
Since $r$ is a global coordinate for $\{\left( \mathcal{O}_{i}, \Psi_{i}\right)\}^{3}_{i=1}$, the previous expression implies that $\dot{r}$ never changes its sign nor it becomes equal to zero. In the case of having with $\gamma$ a future directed curve, the radial coordinate of its points will always be decreasing. This means that for every such a curve, starting from $p = \gamma ( \tau = 0 ) \in S_{r_0}$ into the future, $ r(\tau) \equiv r[\gamma(\tau)]$ will be strictly greater than $r(p) = r_0$ for $\tau > 0$. The disconnected nature of the partition $M_{Sch} = S_{r<r_{0}} \cup S_{r_0} \cup S_{r>r_{0}}$, guarantees then that $I^{+}\left(S_{r_{0}}\right) \subset M_{Sch}\setminus S_{r_{0}}$.


\subsection{Future domain of dependence of $\boldsymbol{S_{r_0}}$: $\boldsymbol{\overline{D^{+}\left(S_{r_{0}}\right)}=S_{r<r_{0}} \cup S_{r_{0}}}$}
\label{subsec:closfutCauchoriz}

Consider a point $p \in S_{r<r_{0}}$ and a smooth timelike curve $\gamma$ that starts from $p$ into the past, that is past inextensible (within $M_{Sch}$) and which has been parameterized by its arc lenght $\tau$. Because of the temporal orientation that has been chosen, the following relation will be satisfied along $\gamma$ (within each $\mathcal{O}_{i}$ containing a segment of it)
	\begin{equation}
	\label{1sttangdecomp}
		\dot{r} = \left(\frac{2M-r}{r}\right)^{1/2} \left[ 1+ \left(\frac{2M-r}{r}\right)\dot{t}^2 + r^2\left(\dot{\vartheta}^2_i + sen^2\vartheta_i\dot{\varphi}^2_i\right) \right]^{1/2}.
	\end{equation}

Under the assumption of the parameter $\tau$ taking on every single value within $[0,\infty)$, it is always possible to give a partition for this interval into segments $[\tau_i,\tau_{i+1})$, characterized by (where $\mathcal{O}_{(i)} \in \{\mathcal{O}_j\}^3_{j=1}$ for every $i$)
	\begin{equation}
	\label{segmntpartitn}
		\begin{gathered}
			\tau_0=0,\\
			\gamma(\tau)\in\mathcal{O}_{(i)}	\quad \forall \tau\in\left[\tau_i,\tau_{i+1}\right),\\
			\gamma(\tau)\in\mathcal{O}_{(i+1)} \quad \forall \tau\in\left[\tau_{i+1},\tau_{(i+1)+1}\right),\; \text{with} \; \gamma(\tau_{\ast})\notin\mathcal{O}_{(i)} \; \text{for some} \; \tau_{\ast}\in \left[\tau_{i+1},\tau_{(i+1)+1}\right).
		\end{gathered}
	\end{equation}

By integrating \eqref{1sttangdecomp} within each one of these intervals (and by taking into consideration the continuity of $r(\tau)$), the inequality $2M > r(\tau) > (2M)^{-1/2} \int^{\tau}_{0} \left(2M-r\right)^{1/2} d\tau'$ follows for every $\tau \in [0,\infty)$. This relation indicates the existence of the limits $\lim_{\tau \rightarrow \infty} \int^{\tau}_{0} \left(2M-r\right)^{1/2} d\tau' $ and $\lim_{\tau \rightarrow \infty} \left(2M-r(\tau)\right)^{1/2}$ \cite{Kudriavtsevbook}, which in turn guarantees that $r(\tau) \longrightarrow 2M$ when $\tau \longrightarrow \infty$. The immediate consequence of this statement is that $\gamma\cap S_{r_{0}}\neq \emptyset$, whenever the arc lenght of such a curve is able to take on every single value within $[0,\infty)$.
	
In order to fully prove the desired identity for $\overline{D^{+}\left(S_{r_{0}}\right)}$, it must also be considered having $\gamma \cap S_{r_0} = \emptyset$ for the curve in question. Nevertheless, the previous procedure forces $\tau$ to be bounded from above for this case. If $\tau_M \equiv \text{sup}\{\tau\}$ so that $\tau \in [0,\tau_M)$, the increasing feature of $r(\tau)$ implies the existence of $r_M\equiv\text{sup}\{r(\tau)\}=\lim_{\tau\rightarrow \tau_{M}} r(\tau)  \leq r_0$. By taking a partition of $[0,\tau_M)$ identical to the one given in \eqref{segmntpartitn}, a similar line of arguments leads now to $2M > r(\tau) > (2M)^{-1}(2M-r_{0}) \int^{\tau}_{0} \vert \dot{t}\vert d\tau'$. Once again, this inequality guarantees that $t_{M}\equiv \lim_{\tau\rightarrow\tau_{M}}t(\tau) \in \mathbb{R}$ \cite{Kudriavtsevbook}.

On the other hand, it is also possible to give the following formulation for $\dot{r}$
	\begin{equation*}
		\dot{r} = \left(\frac{2M-r}{2M}\right)^{1/2} \left[ 1+ \left(\frac{2M-r}{r}\right)\dot{t}^2 + \dot{r}^2 + r^2(\dot{\vartheta}^2_i + sen^2\vartheta_i\dot{\varphi}^2_i) \right]^{1/2}.
	\end{equation*}

The usefulness of this expression resides in the possibility to isolate from it the function $f(\tau) \equiv \{\dot{r}^2(\tau) + r^2(\tau) [\dot{\vartheta}^2_i(\tau) + \sin^2 \vartheta_i (\tau) \dot{\varphi}^2_i (\tau)] \}^{1/2}$. Because of the coordinate transformations between the given atlases, it is obviously the case that this function has in fact a global nature by admitting the formulation $f(\tau) = [ \dot{x}^2(\tau) + \dot{y}^2(\tau) + \dot{z}^2(\tau) ]^{1/2}$. It is then a matter of simple algebra to see that $2M > r(\tau) > (2M)^{-1/2} (2M-r_0)^{1/2} \int^{\tau}_{0} f(\tau')d\tau'$. This again assures the existence of $x_{M} \equiv \lim_{\tau\rightarrow\tau_{M}} x(\tau)$, $y_{M} \equiv \lim_{\tau\rightarrow\tau_{M}} y(\tau)$ and $z_{M} \equiv \lim_{\tau\rightarrow\tau_{M}} z(\tau)$. Since it must be the case that $r_{M}=(x^2_{M} + y^2_{M} + z^2_{M})^{1/2}$ (with $0<r(p)\leq r_{M} \leq r_{0} <2M$), the existence of these three limits, and the one for $t(\tau)$, imply that $\Psi(\gamma(\tau)) = \left( t(\tau), x(\tau), y(\tau), z(\tau) \right) \longrightarrow \left( t_{M}, x_{M}, y_{M}, z_{M} \right) \in \mathbb{R}  \times \mathcal{B}_{o}(2M)$, when $\tau\longrightarrow \tau_{M}$. This obviously contradicts the inextensibility of $\gamma$, and leads to the conclusion that $\gamma \cap S_{r_0} \neq \emptyset$.
	
The content of the previous procedure can be summarized as follows: every smooth timelike curve that starts from $S_{r<r_0}$, and is past inextensible, intersects $S_{r_0}$. This entails that $S_{r_{0}} \cup S_{r<r_{0}} \subset \overline{D^{+}\left( S_{r_{0}} \right)}$ \cite{Waldbook}.

On the other hand, the chosen temporal orientation also implies that $\gamma' \cap S_{r_0} = \emptyset$, for every smooth timelike curve $\gamma'$ that starts from $S_{r>r_0}$ and is past inextensible. By the same reason as before, this last statement implies the inclusion $S_{r>r_{0}} \subset M_{Sch}\setminus \overline{D^{+}\left(S_{r_{0}}\right)}$. The disconnected nature of the partition \eqref{Cauchsurfc} for $M_{Sch}$, indicates then that $\overline{D^{+}\left( S_{r_{0}} \right)} \subset S_{r_{0}} \cup S_{r<r_{0}}$.


\subsection{Future Cauchy horizon of $\boldsymbol{S_{r_0}}$: $\boldsymbol{H^{+}(S_{r_0}) = \emptyset}$}
\label{subsec:Cauchorizn}

For every future directed timelike curve $\gamma$ that starts from some $p \in \overline{D^{+}(S_{r_0})}$, the relation $\gamma \subset S_{r<r_0} \cup S_{r_0}$ always holds. As a consequence of this, $p \in I^{-}\left(\gamma\right) \subset I^{-}\left( S_{r<r_{0}}\cup S_{r_{0}} \right) = I^{-}[\overline{D^{+}\left(S_{r_{0}}\right)}]$ implies the containment $\overline{D^{+}\left(S_{r_{0}}\right)} \subset I^{-}[\overline{D^{+}\left(S_{r_{0}}\right)}]$. The desired result follows from the definition of $H^{+}(S_{r_0})$ as the intersection set $\overline{D^{+}\left(S_{r_{0}}\right)} \cap \{ M_{Sch} \setminus I^{-}[D^{+}\left(S_{r_{0}}\right)] \}$ \cite{Waldbook}.


\subsection{Global hyperbolicity of $\boldsymbol{(M_{Sch},g_{\alpha\beta})}$}
\label{subsec:globhyperb}

An identical procedure to the one given above produces the relations $\overline{D^{-}(S_{r_0})} = S_{r>r_0} \cup S_{r_0}$ and $H^{-}(S_{r_0}) = \emptyset$. By the second one of these, and the result from the previous subsection, the total Cauchy horizon of $S_{r_0}$ equals $\emptyset$ (i.e. $H(S_{r_0}) = H^{-}(S_{r_0}) \cup H^{+}(S_{r_0}) = \emptyset$). In summary, $S_{r_0}$ makes up a closed non-empty set that is not chronological, and for which $H(S_{r_0}) = \emptyset$. Being $M_{Sch}$ connected, all of these properties imply that $S_{r_0}$ forms a Cauchy surface within $M_{Sch}$ \cite{Waldbook}. The spacetime in question is then globally hyperbolic.


\subsection{Closed trapped surfaces}
\label{subsec:trppdsrfcwthoutbndry}

It is a well known fact by now that every 2-sphere of the form $T(t_0,r_0) \equiv \{ p \in M_{Sch} | t(p) = t_0 \; \text{and} \; r(p) = r_0 \}$ (with $t_0 \in \mathbb{R}$ and $r_0 \in (0,2M)$) is actually a closed trapped surface of the whole Schwarzschild spacetime. Nevertheless, it is important to emphasize here what is actually meant by that. A {\it \textup{(}future\textup{)} trapped surface} is a smooth 2-dimensional embedded submanifold $T$ that is spacelike, and whose expansions $\theta_{\pm}$ (of the families of null geodesics departing orthogonally from it into the future) are negative everywhere on $T$. Additionally, a surface like this one is said to be {\it closed} if it is also a compact submanifold without a boundary. If the metric induced on $T$ by $g_{\alpha\beta}$ is represented by $\gamma_{\alpha\beta}$, the numerical value of $\theta_{\pm}$, at each $p \in T$, can be worked out from the formula
	\begin{equation}
	\label{xpansions}
		\theta_{\pm} \equiv \gamma \indices{^{\alpha}_{\beta}} \nabla_{\alpha} k^{\beta}_{\pm} = \sum^{2}_{i=1} \frac{1}{\eta^{\sigma}_{(i)} \eta^{\,}_{(i)\sigma}} \eta^{\alpha}_{(i)} \eta^{\,}_{(i)\beta} \nabla_{\alpha} k^{\beta}_{\pm},
	\end{equation}
where $\{ \eta^{\alpha}_{(i)} \}^{2}_{i = 1}$ is an arbitrary orthogonal basis for the tangent (sub)space to $T$ at $p$.

From the definition of $T(t_0,r_0)$, it follows that the vector fields $(\partial / \partial \vartheta_i)^{\alpha}$ and $(\partial / \partial \varphi_i)^{\alpha}$ generate all the tangent spaces to $T(t_0,r_0) \cap \mathcal{O}_i$. By using these two vectors as the basis $\{ \eta^{\alpha}_{(i)} \}$, and then considering for \eqref{xpansions} the input $
k^{\mu}_{\pm} \equiv f_{\pm}(t_0,r_0) \left[ \left(\partial/\partial r\right)^{\mu} \pm \left(-1 + 2M/r_0\right)^{-1} \left(\partial / \partial t \right)^{\mu} \right]$, the expressions $\theta_{\pm} = 2f_{\pm} (t_0,r_0) / r_0$ are easily derived. The two vectors fields $k^{\mu}_{\pm}$ are future directed for every negative value of $f_{\pm} (t_0, r_0)$. The remaining properties that make each $T(t_0, r_0)$ a closed trapped surface, are of course true.

\bigbreak
Aside from \eqref{Schlinelemnt} satisfying the vacuum Einstein field equations (hence guaranteeing the validity of $R_{\alpha\beta} k^{\alpha} k^{\beta} \geq 0$ for every null $k^{\mu}$), one additional detail must be taken into consideration before drawing upon the theorems to conclude the existence of a singularity. In order to satisfactory identify the existence of a singularity of geodesic incompleteness, the theorems in question require the complementary hypothesis of having an inextensible spacetime. The future inextensibility of $( M_{Sch} , g_{\alpha\beta} )$ follows from the divergent nature of its Kretschmann scalar $R^{\alpha \beta \mu \nu} R_{\alpha \beta \mu \nu} = 48 M^2 / r^6$, when $r \longrightarrow 0^{-}$. Although this last behaviour is frequently used to point out the singularity at $r = 0$, no clear general relationship seems to exist so far between these two kind of singularities (i.e. the ones due to divergent curvature scalars and geodesic incompleteness). Despite some insightful progress has been made in this direction (see \cite{Clarkebook, KannarRacz} and references therein) that kind of scalar divergence will only be used here to guarantee the spacetime inextensibility (nevertheless, a recent proof of the $C^{0}$ inextensibility of the full Schwarzschild-Kruskal spacetime could also be evoked \cite{Sbierski018}). Theorem \ref{penrose65} guarantees then the existence of at least one incomplete, and future directed, null geodesic. As a matter of fact, every radial null geodesic is incomplete.

One last remark must be made before concluding the present section. The previous procedure signals (though in a somewhat measly way) some possibility of using the singularity theorems when analysing spacetimes that, despite being the product of complete gravitational collapse, do not necessarily contain an event horizon.


\section{Schwarzschild spacetime with negative mass}
\label{sec:SchSPnm}

As suggested before, the question of whether or not is possible to rely on this kind of theorems when dealing with spacetimes that contain singularities, but not an event horizon, arises naturally even after analysing the simplest of the examples. The formation of the so called {\it naked singularities}, as a product of gravitational collapses, is nowadays considered a plausible possibility by some authors \cite{Joshibook}. Since the possible existence of singularity theorems applicable to naked singularities is still an open problem to this day (after half a century of the proof of theorems \ref{penrose65} and \ref{hawkpenr70}), a less ambitious approach can be taken. Instead of trying to prove in the most general manner if a (well defined \cite{Waldbook}) naked singularity can be related to the occurrence of geodesic incompleteness, one could try to find out what type of physical information can be gathered about this kind of singularities by means of (the mathematical tools used in) the existing theorems.

The content of the following subsections addresses this issue for the case of the Schwarzschild spacetime with negative mass $M \equiv - m < 0$, and theorems \ref{penrose65} and \ref{hawkpenr70}.


\subsection{The naked singularity Schwarzschild spacetime}
\label{subsec:NSSchwspctm}

The definition of the spacetime to be considered $( M_{SchN} , g^{(N)}_{\alpha\beta} )$ is almost identical to the one given before. The principal differences being the following: (a) The mass parameter $M \equiv -m$ in \eqref{Schwlinelemcarts} is negative. (b) The set $\mathbb{R}^3_o \times \mathbb{R} \subset \mathbb{R}^4$, with $\mathbb{R}^3_o \equiv \mathbb{R}^3 \setminus\{ (0,0,0) \}$, is homeomorphic to $M_{SchN}$ by means of $\Psi$. (c) The open subsets $\mathcal{O}_i$, defined by the maps $\Psi_i$, are given by $\Psi_i ( \mathcal{O}_i ) \equiv (0 , \infty) \times (0,\pi) \times (0,2\pi) \times \mathbb{R}$. (d) The temporal orientation is set by considering as future oriented, all vectors with components $(X^r, X^{\vartheta_i}, X^{\varphi_i}, X^t)$ such that $X^t > 0$. Although these differences are not taken completely into consideration in the upcoming subsections, they are presented here for the sake of completeness. Since all of the next analysis only takes place within one $\mathcal{O}_i$, it is convenient to do $\mathcal{O}_{i} \longleftrightarrow \underline{\mathcal{O}}$, $\Psi_{i} \longleftrightarrow \underline{\Psi}$, $\vartheta_{i} \longleftrightarrow \vartheta$, and $\varphi_{i} \longleftrightarrow \varphi$.


\subsection{No global hyperbolicity}
\label{subsec:nonglobhyperb}

Assuming that $( M_{SchN} , g^{(N)}_{\alpha\beta} )$ is globally hyperbolic, there must exist a {\it global time function} $f$ such that its level hypersurfaces $\Sigma_{f_0} \equiv \{p \in M_{SchN} | f(p) = f_0 \}$ are Cauchy surfaces \cite{Waldbook}. Because of the field $\nabla_{\alpha} f$ being timelike, its temporal component satisfies in $\underline{\mathcal{O}}$ that $\partial_t f \neq 0$. These remarks can be used to see that $| \partial_{r} f / \partial_{t} f | < r / (r+2m)$.

When $f$ is of class $C^1$ in $M_{SchN}$, there exists an open neighbourhood $\mathcal{O}^{(0)} \subset \underline{\mathcal{O}}$ of $p_0 \in \Sigma_{f_0} \cap \underline{\mathcal{O}}$ (for some $f_0$), such that $\widetilde{f} \equiv f - f_0$ (viewed as a function of the tetrads in $\underline{\Psi}(\mathcal{O}^{(0)})$) is of class $C^1$ and such that $(\partial_t \widetilde{f})|_{\underline{\Psi}(p_0)} \neq 0$ and $\widetilde{f}( \underline{\Psi} (p_0)) = 0$. Let $\underline{\Psi} ( p_0 ) \equiv (r_0, \vartheta_0, \varphi_0, t_0)$. In accordance with the implicit function theorem, there exist open sets $J^{(0)} \subset \mathbb{R}$ and $A^{(0)} \subset (0,\infty) \times (0, \pi) \times (0, 2\pi)$ (such that $(r_0, \vartheta_0, \varphi_0, t_0) \in A^{(0)} \times J^{(0)} \subset \underline{ \Psi }( \mathcal{O}^{(0)} )$), that form the image and domain of a function $\widetilde{t}^{(0)} (r, \vartheta, \varphi)$, for which $0 = \widetilde{f}(r, \vartheta, \varphi, \widetilde{t}^{(0)}(r, \vartheta, \varphi))$ for every $(r, \vartheta, \varphi) \in A^{(0)}$. Aside from being $C^{1}$ in its domain, $\widetilde{t}^{(0)}$ satisfies that $\partial \widetilde{t}^{ (0) } / \partial r = - \partial_r f / \partial_t f$. This in turn leads to $| \partial \widetilde{t}^{(0)} / \partial r | < r / (r + 2m)$.

Because of $A^{(0)}$ being open, there exists a maximal curve $\Gamma^{(0)} \equiv \{ \underline{\Psi}^{-1}(r, \vartheta, \varphi, t) | r \in (a_0, b_0), \vartheta = \vartheta_0, \varphi = \varphi_0, t = \widetilde{t}^{(0)} (r, \vartheta, \varphi) \}$ (with $0 \leq a_0 < r_0 < b_0$) that, apart from being $C^1$ and be contained in $\Sigma_{f_0}$, passes through $p_0$ and can be parameterized by its coordinate $r$.

Having a $C^1$ function $t(r) \equiv \widetilde{t}^{(0)} (r, \vartheta_0, \varphi_0)$, whose derivative satisfies $|dt(r) / dr| < r / (r + 2m)$, makes it easier to see the existence of $t_1 \equiv \lim_{r \rightarrow a^+_0} \widetilde{t}^{(0)} (r, \vartheta_0, \varphi_0)$ (as a consequence of the convergence of $\int^{r^{*}}_{r} |dt/dr'|dr'$, with $ 0 \leq a_0 < r^{*} < b_0$, when $r \longrightarrow a^{+}_{0}$). Let $r_1 \equiv a_0$.

Taking a sequence $ \{ q_n \}_{n \in \mathbb{N}} \subset \Gamma^{(0)} $, for which $|r(q_n) - r_1| < 1/n$, produces another sequence $\{ (r(q_n), \vartheta_0, \varphi_0, t(q_n)) \}_{n \in \mathbb{N}}$ that converges to $(r_1, \vartheta_0, \varphi_0, t_1)$. In the case of having $r_1 > 0$, it would also occur that $q_n \longrightarrow p_1 \equiv \underline{\Psi}^{-1}(r_1, \vartheta_0, \varphi_0, t_1) \in \Sigma_{f_{0}} \cap \underline{\mathcal{O}}$, with $r(p_1) = r_1 < r_0 = r(p_0)$, when $n \longrightarrow +\infty$. 

If the same procedure were to be repeated for $p_1$, it would yield another curve $\Gamma^{(1)} \equiv \{ \underline{\Psi}^{-1}(r, \vartheta, \varphi, t) | r \in (a_1,b_1), \vartheta = \vartheta_0, \varphi = \varphi_0, t = \widetilde{t}^{(1)} (r, \vartheta, \varphi) \} \subset \Sigma_{f_0}$ (with $a_1 < r_1 < b_1$, and $\widetilde{t}^{(1)}$ another $C^1$ function) that passes through $p_1$ and can be smoothly matched with $\Gamma^{(0)}$. That is to say, having $r_1 > 0$ would allow to smoothly extend $\Gamma^{(0)}$ through $\Sigma_{f_0}$, to get $\gamma^{(1)} = \Gamma^{(0)} \cup \Gamma^{(1)}$, in a way that increases its domain into $(a_1, b_0)$ (with $a_1 < r_1 < r_0 < b_0$). Once again, if the condition $0 < r_2 \equiv a_1$ would come to happen, it would again be possible to smoothly extend $\gamma^{(1)}$ into a $C^{1}$ curve $\gamma^{(2)}$ with a lower bound $r_3 < r_2$ for its domain. 

This argument can be repeated as many times as a lower radial bound greater than zero is obtained. Since $\Sigma_{f_0}$ is a closed set, this would also be true after an infinite number of steps (or an infinite number of infinite numbers of them), that were to result in a $\gamma^{(\infty)}$ with a new lower bound $r_{\infty} > 0$. Because of this, it is not unrealistic to assume that this construction can only be stopped after obtaining a curve $\gamma$ with $r_{(inf)} \equiv \text{inf}_{p \in \gamma} \{ r(p) \} = 0$. Analogously, it could also be assumed that $\gamma$ does not have a maximum value for its radial coordinate. 

\begin{figure}
		\includegraphics[scale=0.25]{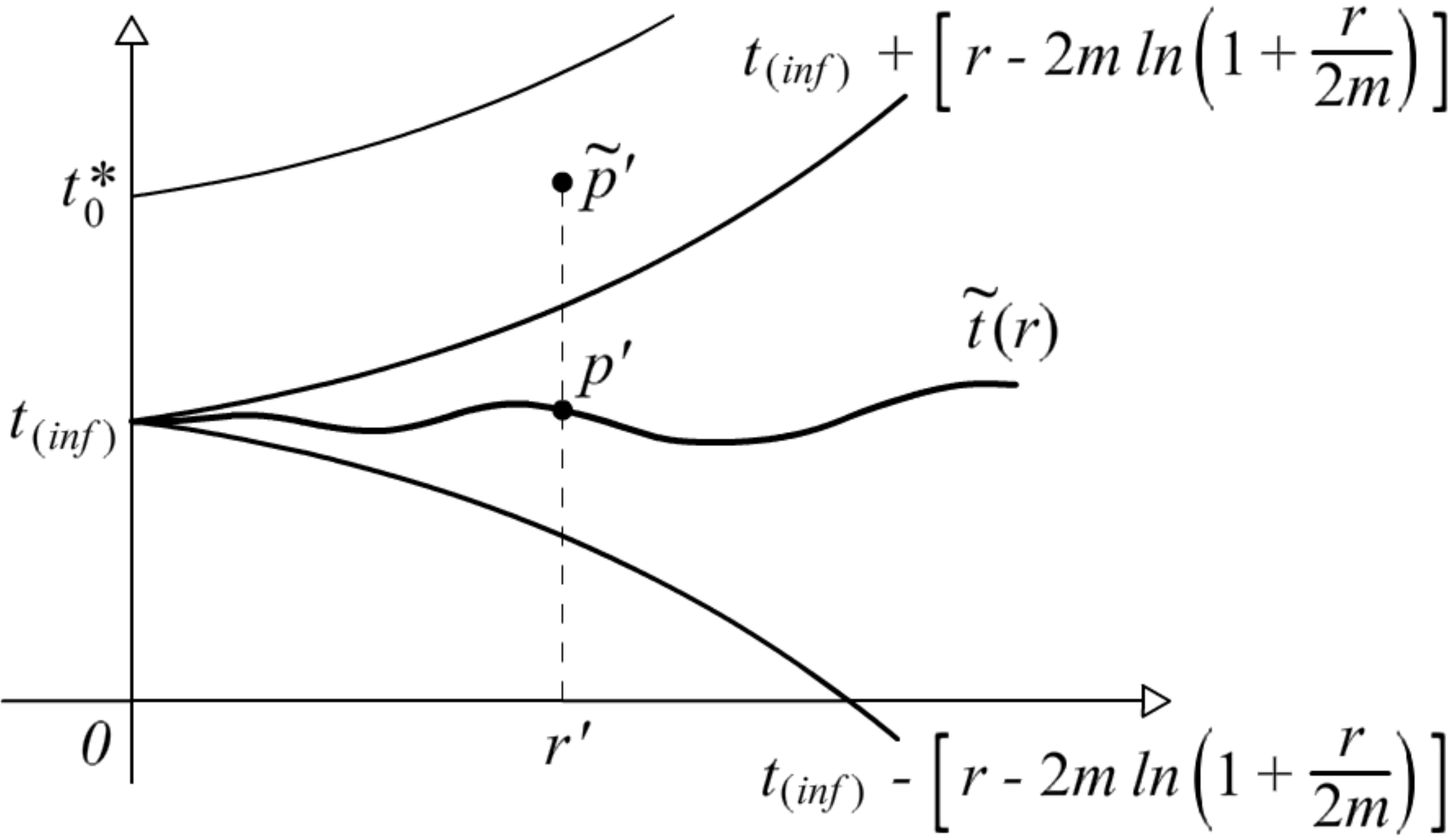}
		\caption{\label{graph.nonglobalhyperb} Inextensible radial null geodesics that do not intersect an assumed Cauchy surface.}
	\end{figure}

All of this reasoning indicates the existence of a curve $\gamma \equiv \{ p \in \Sigma_{f_0} \cap \underline{\mathcal{O}} | r(p) \in (0,\infty), \vartheta(p) = \vartheta_0, \varphi(p) = \varphi_0, t(p) = \widetilde{t}(r(p))\}$, where $\widetilde{t}(r)$ is a $C^{1}$ real function whose derivative satisfies $| d \widetilde{t} (r) / dr | < r / (r + 2m)$. Once again, $t_{(inf)} \equiv \lim_{r \rightarrow 0^+} \widetilde{t}(r) \in \mathbb{R}$. After integrating the last inequality, the following radial behaviour for $\Sigma_{f_0}$ is obtained (Fig.\ref{graph.nonglobalhyperb})
	\begin{equation*}
		t_{(inf)} - \left[ r - 2m \ln\left( 1 + \frac{r}{2m} \right) \right] \leq \widetilde{t}(r) \leq t_{(inf)} + \left[ r - 2m \ln\left( 1 + \frac{r}{2m} \right) \right].		
	\end{equation*}

On the other hand, a solution for the radial and null geodesic equations is given by $t(r) = t^{*}_0 + r - 2m \ln (1 + r/2m)$, with $t^{*}_0$ an arbitrary real constant. Due to the divergent nature of the scalar $R^{\alpha \beta \mu \nu} R_{\alpha \beta \mu \nu} = 48 m^2 /r^6$, every such geodesic is clearly inextensible at $r=0$. Even more so, because of the linear relationship between their coordinate $r$ and affine parameter $\lambda$, all of these geodesics are also incomplete.  

Moreover, every $p\,' \in \gamma$ and $\widetilde{p}\,' \in M_{SchN}$ with $r(\widetilde{p}\,') = r(p\,')$, $\vartheta(\widetilde{p}\,') = \vartheta(p\,')$, $\varphi(\widetilde{p}\,') = \varphi(p\,')$ and $t(\widetilde{p}\,') > t_{(inf)} + r(p\,') - 2m \ln\left[ 1 + r(p\,')/2m \right]$, are chronologically related by $\widetilde{p}\,' \in I^{+}( \Sigma_{f_0})$ (Fig. \ref{graph.nonglobalhyperb}). Bearing this in mind, it is clear that not a single one of this null geodesics will intersect $\Sigma_{f_0}$ if $t^*_0 > t_{(inf)}$. This contradicts the fact that $\Sigma_{f_0}$ is a Cauchy surface \cite{Waldbook}.

\bigbreak
An important remark to make is the following. By knowing all the physical information of the events in a Cauchy surface, it is possible (in principle) to completely determine the (physical) state at every other point in the spacetime. Nevertheless, even if the physical state of every event were to be known at certain global time (represented by an assumed Cauchy surface), the existence of null geodesics that start from $r=0$ at any time would completely ruin total predictability for $(M_{SchN},g^{(N)}_{\alpha\beta})$. Furthermore, it is clear by now that this spacetime is singular at $r=0$, due to the existence of incomplete geodesics that originate from (or end at) that region. Nevertheless, it is not possible to draw this conclusion from theorem \ref{penrose65}.


\subsection{No generic condition}
\label{subsec:nogenercond}

In accordance with theorem \ref{hawkpenr70}, a vector $X^{\alpha}$ is said to be generic if the tensor relation $X_{[\alpha} R_{\beta] \gamma \delta [\mu} X_{\nu]} X^{\gamma} X^{\delta} \neq 0$ holds for it. 

Nevertheless, if $X^{\alpha}$ is not null, this condition turns out to be equivalent to the relation $R_{\alpha \mu \nu \beta} X^{\mu} X^{\nu} \neq 0$ \cite{Beemetal}. Because of this, the tensor identity $R_{\alpha \mu \nu \beta} X^{\mu} X^{\nu} = 0$ can be taken as a system of equations for a non-generic vector $X^{\alpha}$. For the spacetime in question, it is easy to verify that no timelike solution exists. It follows immediately from this that every timelike geodesic of $(M_{SchN}, g^{(N)}_{\alpha\beta})$ is generic at every one of its points. That is to say, any massive object in free fall within $M_{SchN}$ will always experience tidal forces across its points (because of $S_{\alpha \beta} \equiv R_{\alpha \mu \nu \beta} X^{\mu} X^{\nu}$ being the {\it tidal stress tensor} along a geodesic with tangent vector $X^{\alpha}$). In contrast, after imposing the condition $K^{\mu} K^{\nu} g^{(N)}_{\mu \nu} = 0$ on the system of equations $K_{[\alpha} R_{\beta] \gamma \delta [\mu} K_{\nu]} K^{\gamma} K^{\delta} = 0$, the expressions $r^2 (K^r)^2 - (r + 2m)^2 (K^t)^2 = 0$ and $K^{\vartheta} = K^{\varphi} = 0$ end up being the necessary and sufficient conditions for a null vector $K^{\mu}$ to be non-generic. Therefore, no radial and null geodesic can ever be generic in $(M_{SchN}, g^{(N)}_{\alpha \beta})$.

Even though the existence of non-generic geodesics also precludes the use of theorem \ref{hawkpenr70} to guarantee the existence of singularities, insightful information about this spacetime and its singularity can be gained with the help of the mathematical tools used by the theorems.

In order to understand the physical implications of having non-generic null geodesics, it must be taken into consideration the fact that the tensor identity $K_{[\alpha} R_{\beta] \gamma \delta [\mu} K_{\nu]} K^{\gamma} K^{\delta} = 0$, for $K^{\alpha}$ null, is equivalent to the vector map $R\indices{^{\alpha}_{\beta\mu\nu}} K^{\beta} V^{\mu} K^{\nu}$ being proportional to $K^{\alpha}$ every time $V^{\mu} K_{\mu} = 0 $ \cite{Beemetal}. Since this (restricted) map is actually trivial when $K^{\alpha}$ is the tangent vector to any null and radial geodesic in $M_{SchN}$, no tidal acceleration orthogonal to $K^{\alpha}$ will then be experienced (at such geodesics) by any family of geodesics containing them. In particular, no tidal force is ever felt by the family of null geodesics that emerge from any point into the future (or past), along its two radially directed members. 

\begin{figure}
		\subfloat[]{\includegraphics[scale=0.25]{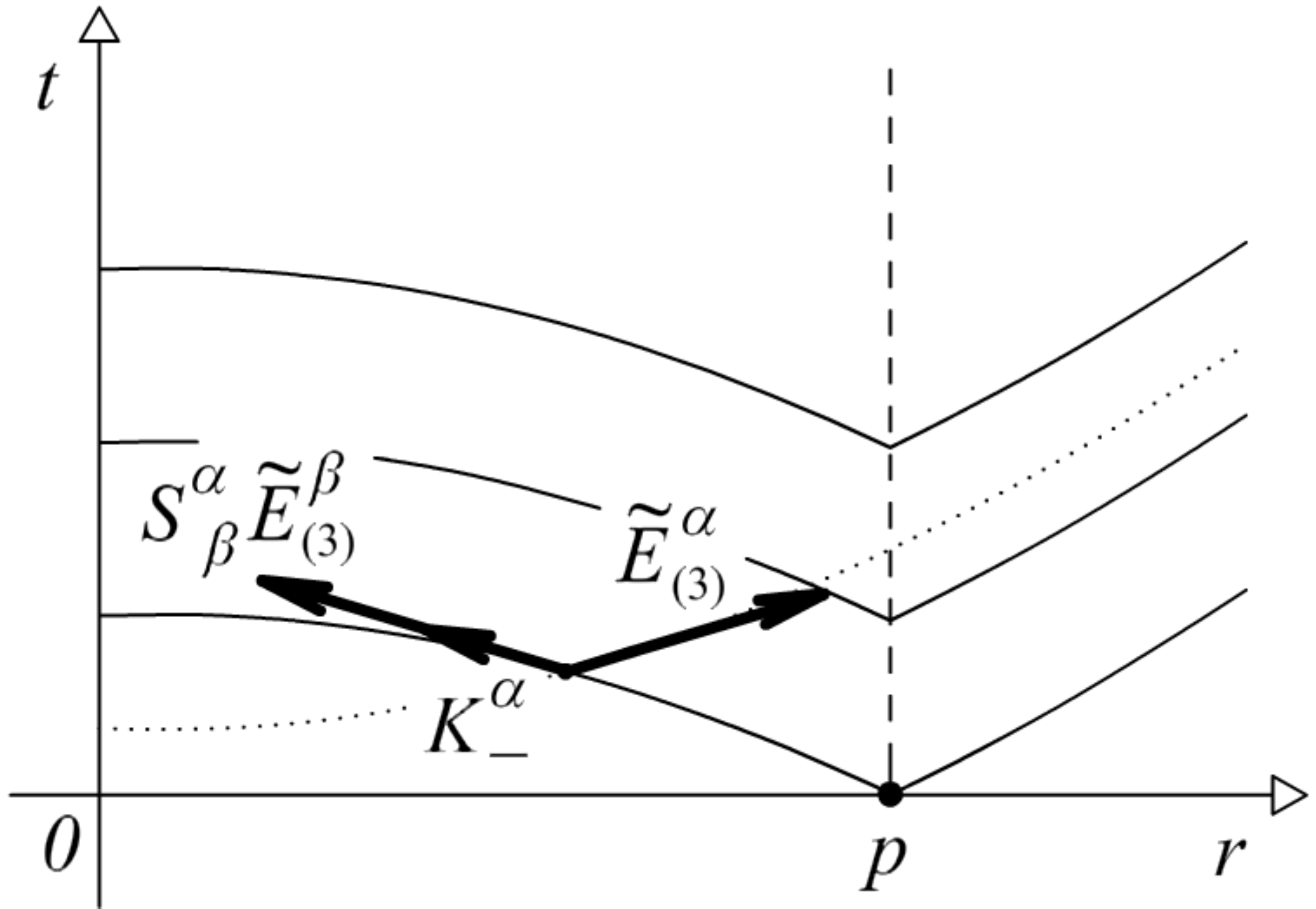}\label{sub.1}} \qquad
		\subfloat[]{\includegraphics[scale=0.26]{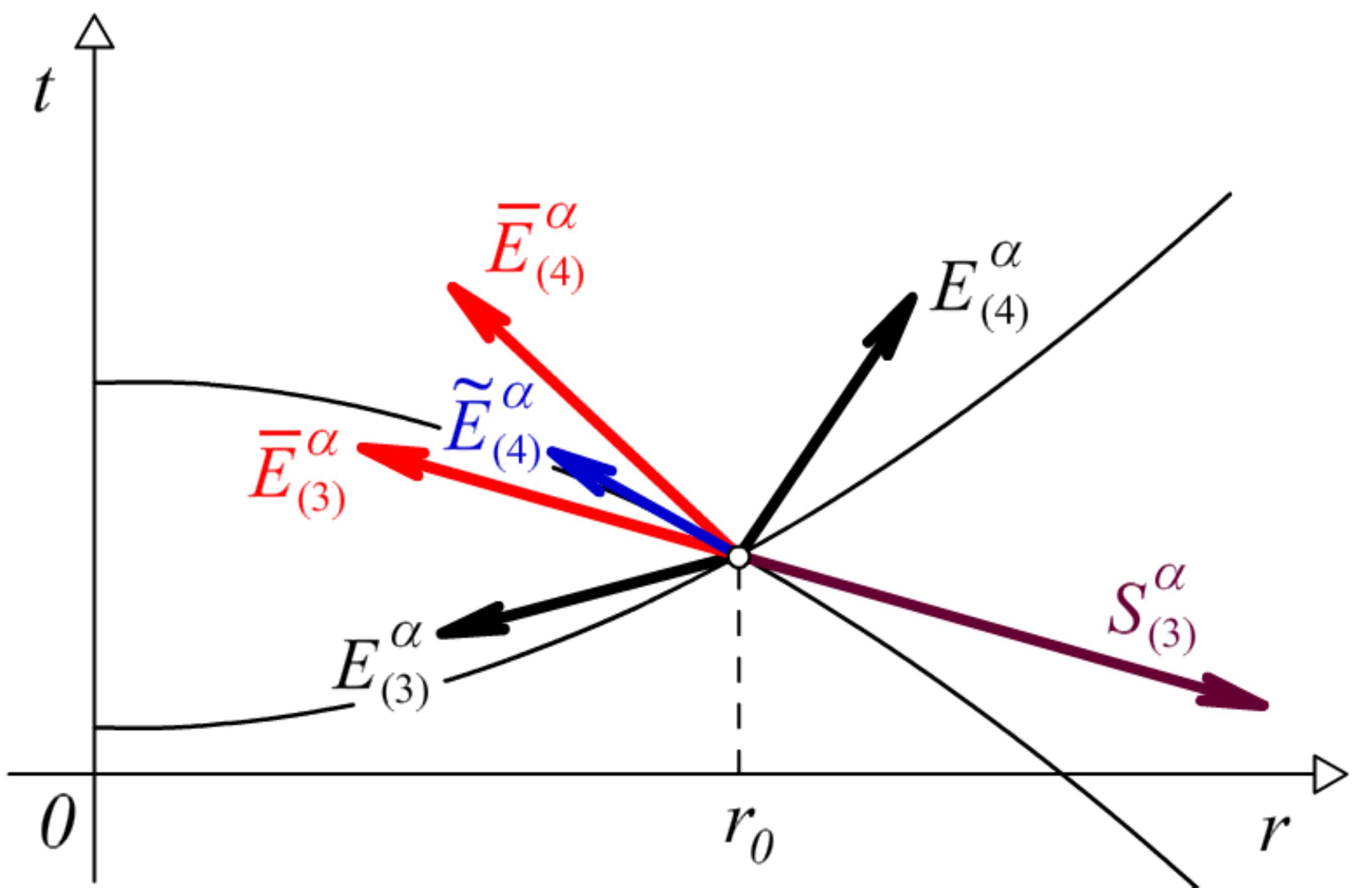}\label{sub.2}}
		\caption{Tidal force vectors.}
		\label{graph.tdlfrc1}
	\end{figure}

This is not to say that no tidal force exists along such geodesics. Let $K^{\alpha}_{-}$ be the tangent vector field to the radial ingoing null geodesics $\gamma_{-}$. If  $\{ \widetilde{E}^{\alpha}_{(i)} \}^{4}_{i = 1}$ is a pseudo-orthonormal basis adapted to $K^{\alpha}_{-}$ (i.e. such that each $\widetilde{E}^{\alpha}_{(i)}$ is parallelly transported along $\gamma_{-}$, with $\widetilde{E}^{\alpha}_{(4)} = K^{\alpha}_{-}$), the corresponding tensor $S\indices{^{ \alpha }_{ \beta }}$ can be decomposed as
	\begin{equation}
	\label{stresstidaldecomp}
		S\indices{^{\alpha}_\beta} \equiv R\indices{^{\alpha}_{\mu \nu \beta}} K^{\mu}_{-} K^{\nu}_{-} = \frac{2m}{r^3} \widetilde{E}^{ \alpha }_{ (4) } \otimes \widetilde{e}^{ (3) }_{ \beta },
	\end{equation}
where $\{ \widetilde{ e }^{(i)}_{\alpha} \}^{4}_{ i = 1 }$ is the corresponding dual basis. Because of this, $S\indices{^{ \alpha }_{ \beta }} \widetilde{E}^{\beta}_{(3)}$ will always point in the direction of $K^{ \alpha }_{ - }$. Since $\widetilde{E}^{\alpha}_{(3)}$ is proportional to the tangent field of the radial outgoing null geodesics, it can be regarded as separating two geodesics $\gamma_{-}$ that depart from the same fixed source at different times (Fig. \ref{sub.1}).  This indicates then an instantaneous relative acceleration towards the singularity for the light waves originating from a continuous fixed source. Even though the previous result can be interpreted as a tendency for light to get compressed towards the singularity (when it has been directly aimed at it), the singularity at $r=0$ is, as will be pointed out later, a timelike one.

On the other hand, for a general incomplete null geodesic with affine parameter $\lambda$, and an adapted pseudo-orthonormal basis $\{ \widetilde{E}^{\alpha}_{(i)} \}^{4}_{i = 1}$, the components $\widetilde{R}_{I 4 4 J} \equiv R_{\alpha \mu \nu \beta} \widetilde{E}^{\alpha}_{(I)} \widetilde{E}^{\mu}_{(4)} \widetilde{E}^{\nu}_{(4)} \widetilde{E}^{\beta}_{(J)}$ (with $I, J = 1, 2$) cannot grow faster in modulus than $|\lambda - \lambda^{*}|^{-2}$ when approaching its singularity at $\lambda^{*}$ \cite{KannarRacz}. This condition is trivially met for every $\gamma_{-}$. Nevertheless, since the radial component of such geodesics takes the form $r(\lambda) = f_{-} (\lambda - \lambda^{*})$ (with $f_{-} < 0$ and $\lambda \leq \lambda^{*}$), this restriction is obviously not satisfied for the remaining components $\widetilde{R}_{\alpha 4 4 \beta}$. This corroborates that the full tensor $S_{\alpha \beta}$ could in fact contain valuable information regarding the divergent nature of the curvature for incomplete null geodesics.

In order to make one last statement with respect to this spacetime not being generic, two additional vector bases must be taken into consideration \cite{Beemetal}. As is easily corroborated, the vector fields
	\begin{equation*}
		\begin{alignedat}{3}
			&E^{\alpha}_{(1)} \equiv \frac{1}{r} \left( \frac{\partial}{\partial \vartheta} \right)^{\alpha} &\qquad & E^{\alpha}_{(3)} \equiv -\frac{r+m}{r} \left( \frac{\partial}{\partial r} \right)^{\alpha} - \frac{m}{r+2m} \left( \frac{\partial}{\partial t} \right)^{\alpha}  \\
			&E^{\alpha}_{(2)} \equiv \frac{1}{r \sin \vartheta} \left( \frac{\partial}{\partial \varphi} \right)^{\alpha} &\qquad &E^{\alpha}_{(4)} \equiv \frac{m}{r} \left( \frac{\partial}{\partial r} \right)^{\alpha} + \frac{r+m}{r+2m} \left( \frac{\partial}{\partial t} \right)^{\alpha},
		\end{alignedat}
	\end{equation*}
are orthonormal (with $E^{\alpha}_{(4)}$ being timelike), parallelly transported along each $\gamma_{-}$, and such that $K^{\alpha}_{-} = -f_{-} [ E^{\alpha}_{(3)} + E^{\alpha}_{(4)}] = f_{-} \left\{ \left( \partial / \partial r \right)^{\alpha} - \left[ r/(r+2m) \right] \left( \partial / \partial t \right)^{\alpha} \right\}$. The second basis is defined by (with $\psi > 0$)
	\begin{equation*}
		\begin{alignedat}{3}
			&\overline{E}^{\alpha}_{(1)} \equiv E^{\alpha}_{(1)} &\qquad & \overline{E}^{\alpha}_{(3)} \equiv \cosh \psi E^{\alpha}_{(3)} + \sinh \psi E^{\alpha}_{(4)} \\
			&\overline{E}^{\alpha}_{(2)} \equiv E^{\alpha}_{(2)} &\qquad &\overline{E}^{\alpha}_{(4)} \equiv \sinh \psi E^{\alpha}_{(3)} + \cosh \psi E^{\alpha}_{(4)},
		\end{alignedat}
	\end{equation*}
and it clearly corresponds to the proper orthonormal basis for an observer moving in the spacial direction of $E^{\alpha}_{(3)}$, at a speed $v = \tanh \psi$, with respect to another one having $E^{\alpha}_{(4)}$ as its 4-velocity. Additionally, despite $\overline{E}^{\alpha}_{(4)}$ being always future directed, it will only point in the direction of $r$ decreasing if (at a given point) $\tanh \psi > m/(r+m)$.

Let $\mathcal{R}_{\alpha \beta \mu \nu}$ and $\overline{R}_{\alpha \beta \mu \nu}$ be the components of the curvature tensor with respect to these bases. Once again, let $I,J =1,2$. Because of $K^{\alpha}_{-}$ being non-generic, it follows that $\widetilde{R}_{I 4 J 4} = 0$ \cite{Beemetal}. This implies having $\overline{R}_{I 4 I 4} = \mathcal{R}_{I 4 I 4} + \left( 1 - e^{-2\psi}  \right) \mathcal{R}_{I4I3}$ and $\overline{R}_{2414} = \mathcal{R}_{2414} + \left[ \left( 1 - e^{-2\psi} \right)/2 \right] \left( \mathcal{R}_{2413} + \mathcal{R}_{2314} \right)$. Additionally, from $\widetilde{R}_{34I4} = 0$ (which is referred to as $K^{\alpha}_{-}$ being {\it non-destructive}) it is straightforward to get $\overline{R}_{3 4 I 4} = e^{-\psi} \mathcal{R}_{34I4}$. It also happens that $\mathcal{R}_{3434} = \overline{R}_{3434} = \widetilde{R}_{3434} = 2m/r^3$. 

Consider now a massive object in free fall from some $p_0$, such that it has $\overline{E}^{\alpha}_{(4)}|_{p_0}$ as its instantaneous 4-velocity. The geodesic deviation equation, together with the non-generic and non-destructive nature of $K^{\alpha}_{-}$, justify the following statement: No significant difference exists between the tidal forces experienced by the points of the object, lying along each direction $\overline{E}^{\alpha}_{(I)}$, and the tidal forces that would be exerted over the object if it were to fall freely from $p_0$ with 4-velocity $E^{\alpha}_{(4)}|_{p_0}$. This assertion is independent of the value given to $\psi$.

On the other hand, perhaps the most interesting conclusion is the one that can be drawn from the remaining curvature component, $\widetilde{R}_{3434}$. According to the geodesic deviation equation, the tidal force associated with the displacement $\bar{J}^{\alpha}|_{p_0} = \bar{J}^{3}_{0} \overline{E}^{\alpha}_{(3)}|_{p_0}$, is given by
	\begin{equation}
		S^{\alpha}|_{p_0} = \bar{J}^{3}_{0} \left[ e^{-\psi} \mathcal{R}_{1434} \overline{E}^{\alpha}_{(1)} + e^{-\psi} \mathcal{R}_{2434} \overline{E}^{\alpha}_{(2)} - \frac{2m}{r^3} \overline{E}^{\alpha}_{(3)} \right]\bigg|_{p_0}.
	\end{equation}

Because of this, if an observer at $p_0$, with 4-velocity $E^{\alpha}_{(4)}$, were to throw a massive object towards the singularity (i.e. along its spacelike direction $E^{\alpha}_{(3)}$) at a speed $v = \tanh\psi > m / \left[r(p_0) + m \right]$, the constituents of the object that lie along its own radial direction $\overline{E}^{\alpha}_{(3)}$ would experience a repulsive tidal force in the opposite direction (Fig. \ref{sub.2}). This force is clearly divergent when $r \longrightarrow 0$. The importance of this feature resides on the fact that $\overline{E}^{\alpha}_{(4)}$ approaches $ \left[ e^{\psi} / \left( 2|f_{-}| \right) \right] K^{\alpha}_{-}$ when $\psi \gg 1$. That is to say, if a massive object were to approach the singularity at $r=0$  by travelling along a near-incomplete-null-geodesic curve, it would have to overcome infinite repulsive tidal forces (aside from the infinite increase in energy necessary to accelerate it to near-light velocities). Even though this repulsive behaviour can already be recognized with the aid of an $r-t$ diagram for the radial timelike geodesics, the preceding argument completely forbids a massive object to ever reach such singularity by travelling along a radial path.

\bigbreak
In summary, the remaining components of \eqref{stresstidaldecomp} not only could provide information regarding the growth of the curvature tensor in a singularity of (null) geodesic incompleteness, they could also be used as indicators of its attractive and/or repulsive nature. A comparison between the divergent behaviour of the stress tidal tensor towards this singularity, and existing extensibility criteria for arbitrary spacetimes (based on H\"older and/or Sobolev norms \cite{Clarkebook}), will also be explored in the future.


\subsection{No trapped points}
\label{subsec:notrppdpnts}

As is well known, equation \eqref{xpansions} can be used to determine the expansion of a null geodesic congruence at anyone of its points. In particular, if a family of affinely parameterized null geodesics $\{\gamma (\lambda) \}$ generate a null hypersurface (so that orthogonal vectors to $K^{\alpha} \equiv \left( \partial / \partial \lambda \right)^{\alpha}$ are associated with trajectories contained in the hypersurface), the same equation can be used to determine the expansion of the generators  $\gamma(\lambda)$ at any point in them. For this particular case, each $\eta^{\alpha}_{(i)}$ in \eqref{xpansions} can be taken as deviation vectors $\left( \partial / \partial s_{i} \right)^{\alpha}$ of one parameter smooth subfamilies $\{ \gamma_{s_i}(\lambda) \} \subset \{\gamma (\lambda)\}$. In view of these facts, the expansion of the geodesics in question can also be reformulated as
	\begin{equation}
	\label{xpansion2}
		\theta = \frac{1}{2} \sum^{2}_{i = 1} \frac{1}{\eta^{\sigma}_{(i)} \eta^{\,}_{(i)\sigma}} K^{\alpha} \nabla_{\alpha} \left[ \eta^{\alpha}_{(i)} \eta^{\,}_{(i)\alpha}  \right],
	\end{equation}
from which it is possible to determine its value along each geodesic. Even more so, in order to get \eqref{xpansion2} for a particular $\gamma_{0}(\lambda) \in \{ \gamma(\lambda) \}$, both vectors $\eta^{\alpha}_{(i)}$ need only be Jacobi fields orthogonal to such geodesics, that satisfy appropriate initial conditions at some $p \in \gamma_{0}(\lambda)$.

Consider now the affinely parameterized family of null geodesics $\{ \gamma^{(0)} (\lambda) \}$ that depart into the future from $p_0 = \underline{\Psi}^{-1} (r_0, \vartheta_0 = \pi/2, \varphi_0, t_0)$. Along the radial ingoing and outgoing geodesics of this family, $\gamma^{(0)}_{-}$ and $\gamma^{(0)}_{+}$, any Jacobi field $\eta^{\alpha}(\lambda)$ would need to be proportional at $p_0$ to the corresponding $\mathcal{K}^{\alpha}_{\pm} \equiv \left( \partial / \partial \lambda \right)^{\alpha}|_{\gamma^{(0)}_{\pm}}$, if it were to be part of a deviation vector field for $\{\gamma (\lambda)\}$. Solving for $\eta^{\alpha}(\lambda)$ the geodesic deviation equation along $\gamma^{(0)}_{\pm}$, the following Jacobi fields can be obtained
	\begin{equation*}
		\eta^{\alpha}_{(1 \pm)} (\lambda) = \eta_{\pm} (\lambda) \left( \frac{\partial}{\partial \vartheta} \right)^{\alpha} \bigg|_{\gamma^{(0)}_{\pm}} \qquad \text{and} \qquad \eta^{\alpha}_{(2 \pm)} (\lambda) = \eta_{\pm} (\lambda) \left( \frac{\partial}{\partial \varphi} \right)^{\alpha} \bigg|_{\gamma^{(0)}_{\pm}},
	\end{equation*}
where $\eta_{\pm}(\lambda) \equiv r_0 (\lambda - \lambda_0)/ \left[ f_{\pm}(\lambda - \lambda_0) + r_0 \right]$ (once again, being $r_{\pm} (\lambda) \equiv f_{\pm} (\lambda - \lambda_0) + r_0$ the radial component of $\gamma^{(0)}_{\pm} (\lambda)$, with sign$(f_{\pm})= \pm 1$). Since $\eta_{\pm} (\lambda_0) = 0$, these two vector fields can be used in \eqref{xpansion2} to calculate the expansion of $\{\gamma^{(0)}(\lambda)\}$ along $\gamma^{(0)}_{\pm}$. The resulting expressions of doing so are $\theta^{(0)}_{\pm} = 2f_{\pm}/ \left[ r_{\pm}(\lambda) - r_0 \right]$, and none of them ever become negative when traversing the corresponding geodesics into the future. Because of the spherical symmetry of this spacetime, the aforementioned result for $\gamma^{(0)}_{+}$ indicates that no trapped point exists within $(M_{SchN},g^{(N)}_{\alpha\beta})$.

Two important conclusions can be drawn from the previous results. First, since a family of null geodesics departing from a point into the future represents a pulse of light being emitted in all directions during a single event, the lack of trapped points in $M_{SchN}$ can effectively be viewed as the absence of regions of inescapable confinement. This result is in agreement with the existence of incomplete null geodesics that start from the singularity, and can be extended indefinitely into the future. Second, the limit $\theta^{(0)}_{-} \longrightarrow 2|f_{-}|/r_0 \in \mathbb{R}$, when $r_{-}(\lambda) \longrightarrow 0^{-}$, is in agreement with the previously asserted lack of tidal forces experienced by $\{\gamma^{(0)} (\lambda)\}$ along $\gamma^{(0)}_{\pm}$. According to this, the null geodesics from an arbitrary point into the future even experience an unaccelerated expansion towards the singularity at $r=0$. This means that even the nearly radial geodesics in $\{ \gamma^{(0)} (\lambda) \}$ will avoid the singularity in question. As was previously anticipated, this property emphasizes its temporal nature.

 
\subsection{Trapped surfaces}
\label{subsec:trppdsrfcs}

It has been proven by now (in a rather elegant way) that closed trapped surfaces cannot exist within stationary spacetimes \cite{MarsSenovilla}. In order to understand the meaning and implications of the possible existence of trapped surfaces that are not necessarily closed, a quick review of the main tools used in such a proof is needed. For the sake of simplicity, the arguments presented next will be restricted to $(M_{SchN}, g^{(N)}_{\alpha\beta})$. However, the generalizations of the formulae and concepts that do not specifically refer to quantities of this spacetime, are plainly true.

Consider a 2-dimensional smooth manifold $\Sigma$ that is orientable, an embedding from it into $M_{SchN}$, $\Phi : \Sigma \longrightarrow S \equiv \Phi(\Sigma) \subset M_{SchN}$, and a $C^{1}$ vector field $\xi^{\alpha}$ (defined within an open neighbourhood of $S$) that generates a one parameter group of diffeomorphisms $\{\phi_{\tau}\}_{\tau \in (a,b)}$, such that $\phi_{\tau = 0 (\in (a,b))}$ is the identity map. Additionally, assume that a family of metrics on $\Sigma$ can be defined by means of the quantities $\gamma_{AB} (\tau) \equiv \{ g^{(N)}_{\alpha \beta} \left[ \Phi^{*}_{\tau} e_{(A)} \right]^{\alpha} \left[ \Phi^{*}_{\tau} e_{(B)} \right]^{\beta} \}|_{S_{\tau}}$, where each $e_{(A)}$ ($A=1,2$) is a basis vector field for $\Sigma$, and $\Phi^{*}_{\tau}$ is the push-forward map associated with the embedding $\Phi_{\tau} \equiv \phi_{\tau} \circ \Phi: \Sigma \longrightarrow S_{\tau} \equiv \Phi_{\tau}(\Sigma) \subset M_{SchN}$. From its definition, the matrix $(\gamma_{AB}(\tau))$ also comprises the components of a tensor $\gamma_{\alpha\beta}(\tau)$ that equals the metric induced on $S_{\tau}$ by $g^{(N)}_{\alpha \beta}$, when $\text{det} (\gamma_{AB}(\tau)) \neq 0$.

If $\gamma_{\alpha \beta} \equiv \gamma_{\alpha \beta}(\tau = 0)$ is positive definite, two future directed continuous null vector fields $k^{\alpha}_{\pm}$, orthogonal to $S$, can be constructed. By using them, it is straightforward to get the following identity for the (instantaneous) variation of the induced area elements $\boldsymbol{\eta}_{\Sigma} (\tau)$
	\begin{equation}
	\label{volvar}
		\frac{d \boldsymbol{\eta}_{\Sigma} (\tau)}{d \tau} \bigg|_{\tau = 0} = \frac{1}{2} \left[ \! \mbox{\large $\mathsterling$}_{\xi}g^{(N)} \right]_{\alpha \beta} \gamma^{\alpha \beta} \boldsymbol{\eta}_{\Sigma} = \left( \text{div}_{S}\xi_{\parallel} + \xi_{\mu} H^{\mu} \right) \boldsymbol{\eta}_{\Sigma}.
	\end{equation} 
where $\text{div}_{S} \xi_{\parallel} \equiv \gamma\indices{^{\alpha}_{\beta}} \nabla_{\alpha} (\gamma\indices{^{\beta}_{\mu}} \xi^{\mu})$ is the induced divergence of the parallel projection of $\xi^{\alpha}$ on $S$, and $H^{\alpha} \equiv \left( k^{\mu}_{+} k^{\,}_{-\mu} \right)^{-1} \left( \theta_{-}k^{\alpha}_{+} + \theta_{+}k^{\alpha}_{-} \right)$ (with $\theta_{\pm}$ given by \eqref{xpansions}) is the so called {\it mean curvature vector field} of $S$. The advantage of introducing $H^{\mu}$ resides in the fact that $S$ being a future trapped surface is equivalent to $H^{\mu}$ being timelike and future directed. 

According to the second expression of \eqref{volvar}, if $\xi^{\alpha}$ were to be replaced by a smooth extension of one of the fields $k^{\alpha}_{\pm}$, the term between parentheses would reduce to $\theta_{\pm}$. This result justifies the physical interpretation given in section \ref{sec:roST} for a trapped surface $S_{t}$. Nevertheless, it also follows from \eqref{volvar} that even if such a surface were to exists within $M_{SchN}$, no inescapable confinement scenario would be hinted by it. This assertion is founded by the fact that, for a future directed timelike Killing vector field $\xi^{\alpha}$, the previous equation would entail for $S_{t}$ the relation $0 < -\xi_{\mu} H^{\mu} = \text{div}_{S_{t}} \xi_{\parallel}$. Since no such $S_{t}$ can then have $\xi^{\alpha}_{\parallel} \equiv \gamma\indices{^{\alpha}_{\beta}} \xi^{\beta} \equiv 0$, no simultaneous emission of two shrinking area light beams (relative to the coordinate time of the -global- temporal translation symmetry) could be represented by it (contrary to every $T(t_0,r_0)$ of section \ref{subsec:trppdsrfcwthoutbndry}, with respect to the global time $r$ of $(M_{Sch},g_{\alpha\beta})$). 

That is not to say that trapped surfaces could not be of any significance for stationary spacetimes. In general, the induced divergence of a non vanishing vector field $\zeta^{\alpha}$, tangent to $S_t$, can be expressed as $\text{div}_{S_{t}} \zeta = \gamma^{-1/2} ( \partial \gamma^{1/2} / \partial v )$, where $\gamma \equiv \text{det} (\gamma_{AB} (0))$ and $(\partial / \partial v)^{\alpha} = \zeta^{\alpha}$ for some right handed coordinate system on $S_{t}$. According to this, the inequality $\text{div}_{S_{t}} \xi_{\parallel} > 0$ would in fact imply the existence of a direction field over $S_t$, $-\xi^{\alpha}_{\parallel}$, along which the induced area element would always decrease. Because the divergence in question is a scalar, this conclusion is coordinate invariant.

On the other hand, the aforementioned proof of \cite{MarsSenovilla} strongly bases its main argument on the impossibility of having trapped surfaces that, simultaneously, be compact and lack a boundary (as embedded submanifolds).This fact alone does not prevent the existence of mere trapped surfaces, so the orientability and compactness of such possible surfaces could be referred to as them being {\it physically feasible}. That is to say, any physically feasible trapped surface within a stationary spacetime must have a boundary.

As a matter of fact, the existence of (non-compact) trapped surfaces even within Minkowski spacetime has been established for quite a while by now \cite{Senovill1}. This clearly opens up the possibility of finding non-closed trapped surfaces in the (static) spacetime $(M_{SchN}, g^{(N)}_{\alpha \beta})$. In particular, inspired by the shape of such examples, several non-compact trapped surfaces can be constructed. Perhaps the most illustrative one is the 2-dimensional submanifold $S_{T}$, defined as the intersection of two hypersurfaces given by
	\begin{equation}
	\left\{
		\begin{gathered}
			t + \frac{1}{2} \left[ r - 2m \ln\left( 1 + \frac{r}{2m} \right) \right] = 0, \\
			z = 0.
		\end{gathered}
	\right.
	\label{trappdsurfSchN}
	\end{equation}

\begin{figure}
		\includegraphics[scale=0.3]{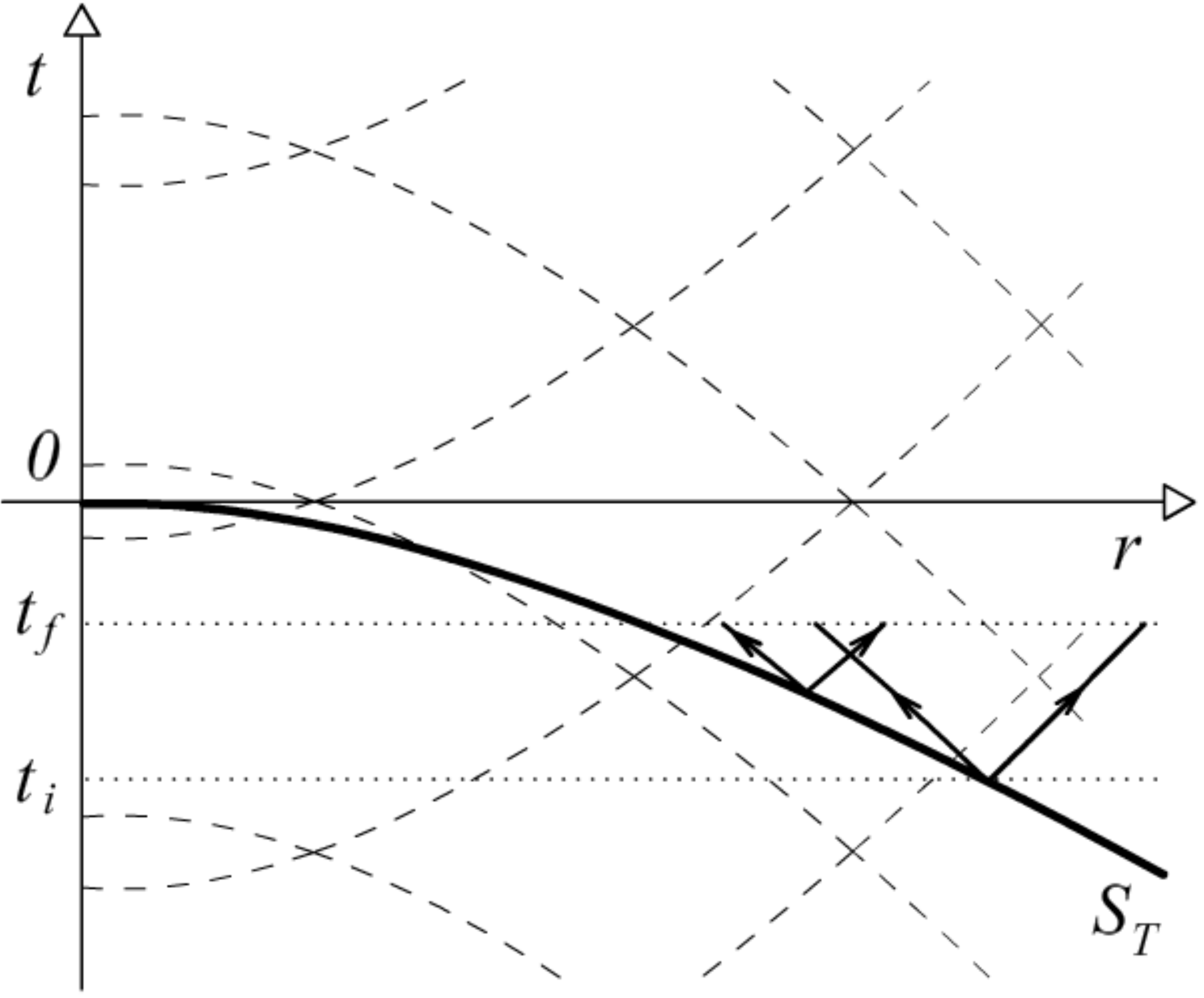}
		\caption{\label{graph.SchNtrpsrfc} Non-compact trapped surface $S_{T} \subset M_{SchN}$.}
	\end{figure}

As can easily be verified, $S_{T}$ turns out to be a spacelike submanifold without a boundary that, by having  $\theta_{\pm} = -2(r + m) / (3r^2)$, forms a non-compact orientable trapped surface (Fig. \ref{graph.SchNtrpsrfc}). Naturally, physically feasible trapped surfaces can then be obtained by extracting smooth compact portions from it (guaranteeing in this way the possibility to smoothly extend their orthogonal fields $k^{\alpha}_{\pm}$, and expansions $\theta_{\pm}$, to their boundary points). To construct such feasible surfaces, families of light rays must be emitted simultaneously from points along rings of constant $r$ (contained in the equatorial plane) at different times $t(r)$ given by \eqref{trappdsurfSchN} (Fig. \ref{graph.SchNtrpsrfc}). Once again, no inescapable confinement is represented then by $S_{T}$.

Finally, for the parallel projection of $\xi^{\alpha} = (\partial / \partial t)^{\alpha}$ on $S_{T}$ (given by $\xi^{\alpha}_{\parallel} = [2(r + 2m) / (3r)] \left( \partial / \partial r \right)^{\alpha} - (1/3) \left( \partial / \partial t \right)^{\alpha}$), the induced divergence is $\text{div}_{S_{T}} \xi_{\parallel} = 2(r+m) / (3r^2)$. In accordance with the previous remarks, $-\xi^{\alpha}_{\parallel}$ points in the direction towards which the induced area element decreases. Even more so, since $\text{div}_{S_{T}} \xi_{\parallel} \longrightarrow +\infty$ when $r \longrightarrow 0^{-}$, $-\xi^{\alpha}_{\parallel}$ corresponds in fact to a direction of infinite area contraction along the non-compact trapped surface $S_T$. 

Although a similar divergent behaviour occurs for the trapped surface (within Minkowski spacetime) of \cite{Senovill1}, an infinite extent must be traversed for it to be reached. Because in $(M_{SchN},g^{(N)}_{\alpha \beta})$ such contraction coincides with the presence of a scalar/geodesic-incompleteness singularity, it is suggested here that any extreme behaviour of $\text{div}_{S_t} \xi_{\parallel}$ could in fact be used as an indicator of the existence of intrinsic ``pathological regions'' in a (stationary) spacetime. That is to say, the existence of (non-closed) trapped surfaces could still be used to hint the presence of singularities, even if they do not correspond to situations of inescapable confinement.

\bigbreak
On the other hand, it is important to mention that hypothesis $(4) (\text{a})$ of theorem \ref{hawkpenr70} has not been taken into consideration due to the fact that its usual physical interpretation allures only to ``closed'' universes \cite{HawkElli}. Nevertheless, its physical implications will be discussed in future works for the sake of completeness, together with an analogous analysis for the trapped submanifolds of co-dimension 3 (i.e. trapped curves) of \cite{GallowaySenovill}.


\section{Conclusions}
\label{sec:conclus}

This work presents a complete usage of the classical singularity theorems of Penrose (1965) and Hawking-Penrose (1970). Together with appropriate and illustrative interpretations for all of their hypothesis, the applicability of the theorems themselves or the mathematical tools used in them is discussed for the Schwarzschild spacetimes (i.e. the ones with positive and negative mass).

A rigorous treatment for the positive mass case from the point of view of Penrose's theorem (that seemingly does not exist so far in the literature), is discussed in detail. As is expected from the way a spacetime is handled by the theorems, the sole definition of the Schwarzschild spacetime is enough to guarantee the existence of a singularity by such theorem. The main result for this case is the validity of the hypotheses of the theorem in a spacetime that, despite being part of a black hole (i.e. the possible outcome of a process of complete gravitational collapse), does not contain an event horizon.

For the second case, it is showed the impossibility of using either of the theorems to assure the existence of singularities. From the proof of such spacetime not being globally hyperbolic, two conclusions are drawn: (1) there does indeed exist incomplete null geodesics within this spacetime, and (2) their presence is responsible for the absence of total predictability.

Additionally, it is showed that even though the radial null geodesics of this spacetime are all non-generic, the temporal nature of their singularity can be better understood from this fact. Even more so, is it presented with this analysis an example of how the full components of the tidal stress tensor along such null geodesics, with respect to pseudo-orthonormal bases adapted to them, could in fact provide information regarding divergent curvature behaviour in the singularity, as well as its possible repulsive or attractive nature. 

In regard to this singularity being naked, agreement is found between the existence of incomplete null geodesics that extend to infinity, and the absence of regions of inescapable confinement suggested by a lack of trapped points (and closed trapped surfaces). Furthermore, the occurrence of non-closed trapped surfaces within this spacetime is discussed. It is argued how despite them not representing scenarios of inevitable confinement, they still can suggest the existence of singularities in static spacetimes. 

Further analysis for these features, in cases of more general static (or even stationary) spacetimes, will be explored in the future.


\section*{Acknowledgements}

This work was partially supported  by UNAM-DGAPA-PAPIIT, Grant No. 114520, 
Conacyt-Mexico, Grant No. A1-S-31269, and by the Ministry of Education and Science (MES) of the Republic of Kazakhstan (RK), Grant No. 
BR05236322 and AP05133630.






\end{document}